# Worm-Like-Chain Model of Forced Desorption of a Polymer Adsorbed on an Attractive Wall

Pui-Man Lam[*] and Yi Zhen[+]
Physics Department, Southern University
Baton Rouge, Louisiana 70813

Forced desorption of a semiflexible polymer chain on a solid substrate is theoretically investigated. The pulling force versus displacement is studied for different adsorption energy $e$ and persistence length P. It is found that the relationships between pulling force and cantilever displacement show a series of characteristic force spikes at different persistence length P. These force spikes becomes more pronounced but the average magnitude of this force decrease as P grows. Our results are of relevance to forced desorption of DNA on an attractive wall in single-molecule pulling experiments

* puiman_lam@subr.edu ;  + yi_zhen@subr.edu

**I. Introduction**

In the past decade, force has been used as a thermodynamic variable to understand the elastic, mechanical, structural, and functional properties of biomolecules[1-3]. The dependence of force provides basic understanding of the interactions such as end-to-end distance versus force which can be used to follow the progress of the interaction[3]. Experimentally, it has become possible to directly measure the forced desorption of a single polymer molecule in contact with a solid surface[4-12]. In these experiments, forced desorption process is realized by single-molecule pulling technique through exerting a force in pN range. The single polymer molecules are chemically attached to an atomic force microscope(AFM) tip. The attached polymers are then brought into contact with and removed from a solid substrate subsequently. During this process, the force required to constrain the polymer at a given height above the substrate is measured and force-extension relationship is studied. Theoretically, many efforts have been made to interpret experimental observations of forced desorption of polymers using equilibrium model within a master-equation approach for the case of constant velocity in a AFM experiment, Bell-Evans equation which provides mean detachment force as a function of temperature T and loading rate, and Brownian dynamics simulation for forced-induced desorption under force control and displacement control[13-17].

Experiments of desorption of polymers generally show a rapid increase in the force for small distance between substrate and cantilever and an extended force plateau as the molecule is gradually peeled off the substrate. The theoretical analyses of the desorption of polymer have been carried out using models such as freely jointed chain (FJC) or wormlike chain model (WLC)[18-20]. In WLC, the persistence length is introduced. Persistence length is an important parameter for semiflexible polymers such as DNA which has the persistence length of the most prominent biopolymers and ranges from 50nm. On scales which exceed persistence length P, the orientational order of the polymer segments decays exponentially and the polymer effectively behaves as a flexible chain with segment size set by P. In contrast, on length scales which are small compared to P, the bending energy of the semiflexible polymer plays an equally important role and strongly affects the behavior of the polymer. Semiflexibility is crucial factor for the adsorption onto adhesive substrate.[21]. Desorption transitions arise from the competition between the energy gained by binding to an attractive potential wall and the associated loss of configuration entropy. The entropy loss is reduced and adsorbed more easily with increasing persistence length.

Recently Paturej et al [22] study the desorption of a polymer adsorbed on an attractive wall by pulling with an external force at one end of the polymer. In their study they have used the FJC model. The present paper is devoted to readdressing the same problem using WLC model. With the WLC model we can study the effect of the persistence length on the desorption. Our model is relevant to forced desorption of DNA from an attractive wall. The persistent length of DNA can be varied by interaction with protein in solution [23-26].

The organization of the paper is as follows: In Sec. II the general free energy functions depending on coarse-grained variables is constructed in presence of the external force. The equilibrium theory of detachment for the case of strong polymer adsorption is investigated and the mean force displacement relationships are studied in detail. Sec.III is the conclusion of what have been obtained and a discussion of future direction.

## II. Model and Results

Recently Paturej et al [22] study the desorption of a polymer adsorbed on an attractive wall by pulling with an external force at one end of the polymer. They have used the FJC model to describe the polymer. Here we readdress the same problem, but replacing the FJC model by the WLC model. The WLC model describes a semi-flexible chain with an extra parameter P, the persistent length of the polymer. This model is more appropriate to describe double stranded DNA. Using the model we can study the effect of the persistent length on forced desorption of DNA. For double stranded DNA the persistent length is about 50nm. But this value can be varied using interaction with protein in solution [23-26]. Figure 1 illustrates the principal scheme of a single molecule forced desorption experiment based on AFM. R denotes the distance between the clamped end of the desorbed portion and the substrate and D denotes the distance between the cantilever and the substrate. N-n and n denote the number of adsorbed and desorbed monomers respectively. The total partition function for a fixed cantilever distance D i.e. $Z_{tot}(D)$ is a product of the partition functions of the adsorbed part, $Z_{ads}(n)$ of the adsorbed portion, of the desorbed portion (a stretched polymer portion), $Z_{pol}(n,R)$ and of the cantilever itself $Z_{can}(D-R)$. As a result

$$Z_{tot} = \sum_{n=0}^{N} Z_{ads}(n) \int_0^{bn} dR Z_{pol}(n,R) \boldsymbol{q}(D-R) Z_{can}(D-R) , \qquad (1)$$

where N is the total number of monomers in the polymer and b is the segment length of one monomer. By introducing the function $\min(bn,D) = bn$, if bn<D and $\min(bn,D) = D$, if D<bn, eqn. (1) can be written as

$$Z_{tot} = \sum_{n=0}^{N} Z_{ads}(n) \int_0^{\min(bn,D)} dR Z_{pol}(n,R) Z_{can}(D-R) \qquad (2)$$

In the strong interaction regime, $Z_{ads}(n)$ attains the simple form

$$Z_{ads}(n) = \exp(\tilde{e}(N-n)),  \qquad (3)$$

where $\tilde{e} = \dfrac{e}{k_B T}$ is the dimensionless adsorption energy with T the absolute temperature and $k_B$ is the Boltzman constant. The cantilever manifests itself as a harmonic spring with spring constant $k_c$, with the corresponding partition function

$$Z_{can}(D-R) = \exp\left[-\dfrac{k_c}{2k_B T}(D-R)^2\right]. \qquad (4)$$

Finally we derive the partition function for the desorbed part of the polymer. In the WLC model, the force-extension curve is given by [27]

$$\tilde{f}(R) = \dfrac{b}{P}\left[\dfrac{1}{4}\dfrac{1}{\left(1-\dfrac{R}{nb}\right)^2} - \dfrac{1}{4} + \dfrac{R}{nb}\right] \qquad (5)$$

where the dimensionless force is defined as $\tilde{f} = \dfrac{fb}{k_B T}$ and P is the persistence length. This is an interpretation formula proposed by Marko and Siggia [27] which reproduces the experimental result very well.

The work done in stretching the polymer to a distance R is given by

$$W = \int_0^R \dfrac{k_B T}{b}\tilde{f}(R')dR'. \qquad (6)$$

Using Eqn. (5), this integral can be easily calculated as

$$W = -nk_B T G(R)  \qquad (7)$$

with $G(R)$ defined as

$$G(R) = \dfrac{b}{P}\left\{\dfrac{1}{4}\left(\dfrac{R}{nb}\right)\left[1-2\left(\dfrac{R}{nb}\right)\right] - \dfrac{1}{4}\left[1-\left(\dfrac{R}{nb}\right)\right]^{-1}\right\}. \qquad (8)$$

From this we can write the partition function for the desorbed part of the polymer as

$$Z_{pol}(n,R) = \exp\left[-\dfrac{W}{k_B T}\right] = \exp[nG(R)]. \qquad (9)$$

Substituting this and Eqns. (3) and (4) in Eqn.(2), the total partition function now reads

$$Z_{tot} = \sum_{n=0}^{N} \exp(\tilde{e}(N-n)) \int_{0}^{\min(bn,D)} dR \exp(nG(R)) \exp\left[-\frac{k_c}{2k_BT}(D-R)^2\right]. \quad (10)$$

The average force $<f_z>$ measured by AFM experiment is given by

$$<f_z> = -k_BT \frac{\partial}{\partial D} \ln Z_{tot} = \frac{k_c}{Z_{tot}} \sum_{n=0}^{N} \exp(\tilde{e}(N-n)) \int_{0}^{\min(bn,D)} dR(D-R) \exp(nG(R)) \exp\left[-\frac{k_c}{2k_BT}(D-R)^2\right]$$
(11)

In Figure 2 we show the average force versus cantilever distance D, for two values of the cantilever spring constant $k_c = 10$, and $k_c = 100$, for values of the persistent length P=1, P=5 and P=10. The adsorption energies are kept at $\tilde{e}=2$, $\tilde{e}=5$ and $\tilde{e}=16$ in figures (2a), (2b) and (2c) respectively. All lengths are measured in units of the monomer length b. From these figures one can see that the persistent length P has an effect on the average force. The average force verses cantilever displacement curves show a series of characteristic force peaks. These peaks become more pronounced as the persistent length P increases but the mean value seems to decrease with P. This is consistent with the reduced entropy loss with increasing persistent length and makes the desorption easier. As with the case of the FJC, the spikes correspond to the reversible transition $n \leftrightarrow n+1$, during which the release of polymer stretching energy is balanced by adsorption energy.

In order to compare with the FJC model we have reproduced the results of that model for the same values of the adsorption energies and spring constants in Figures (3a)-(3c). Comparing Figures (2) and (3) we can see that our WLC results are closest to the FJC results for persistent length P=1, consistent with the fact that the FJC is a flexible polymer with vanishing persistent length.

### III. Conclusions

We have investigated the forced desorption of a polymer adsorbed on an attractive wall using the WLC model. Our results are similar to those obtained using the FJC model, especially at very small values of the persistent length P. The average force versus cantilever displacement show a series of characteristic force peaks. These peaks become more pronounced as the persistent length P increases but the mean value seems to decrease with P. This is consistent with the reduced entropy loss with increasing persistent length and makes the desorption easier. As with the case of the FJC, the spikes correspond to the reversible transition $n \leftrightarrow n+1$, during which the release of polymer stretching energy is balanced by adsorption energy.

Since the WLC model can better describe DNA stretching, our results are of relevance to forced desorption of DNA adsorbed on an attractive wall in single-molecule experiments. The persistent length of double-stranded DNA is about 50nm, but this value can be varied using interaction of protein with DNA in solution [23-26].

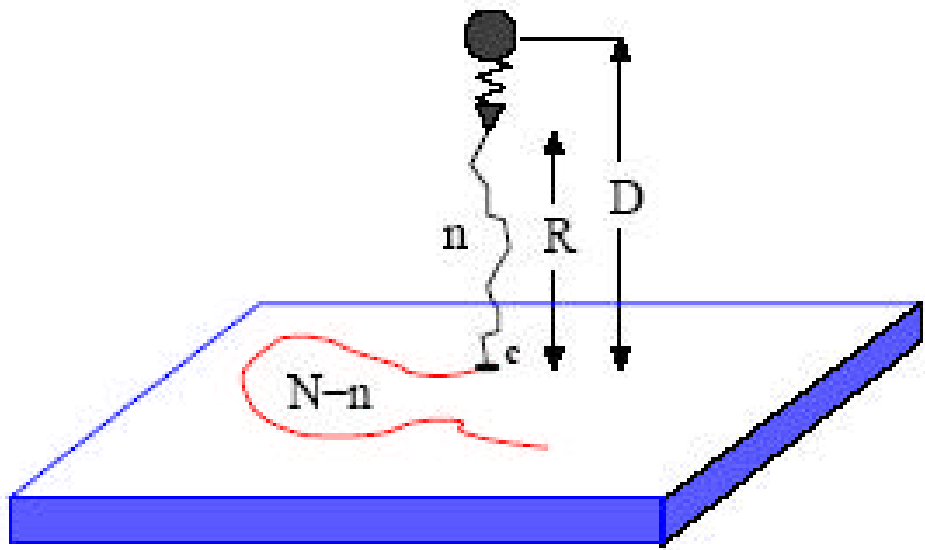

Figure 1: Principal scheme of a single molecule forced desorption based on the AFM. R denotes the distance between the clamped end of the desorbed portion and the substrate and D denotes the distance between the cantilever and the substrate. N-n and n denote the number of adsorbed and desorbed monomers respectively.

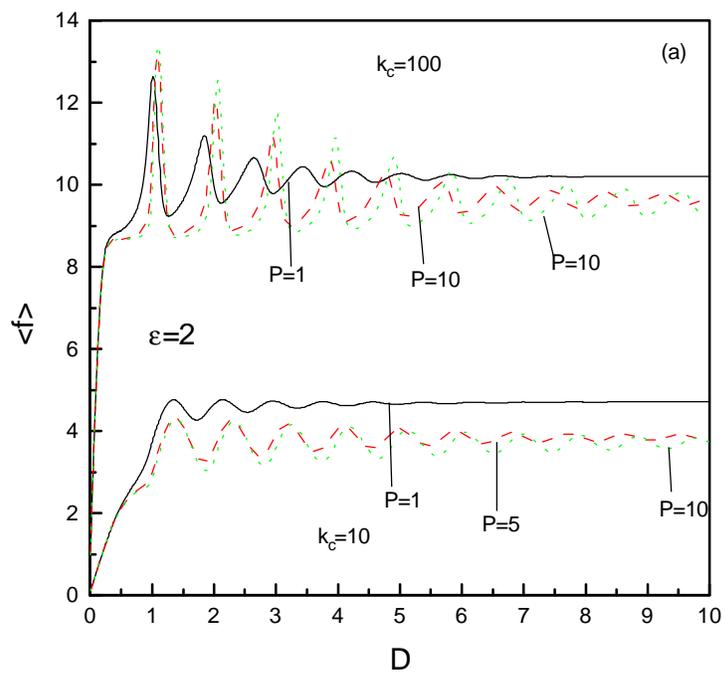

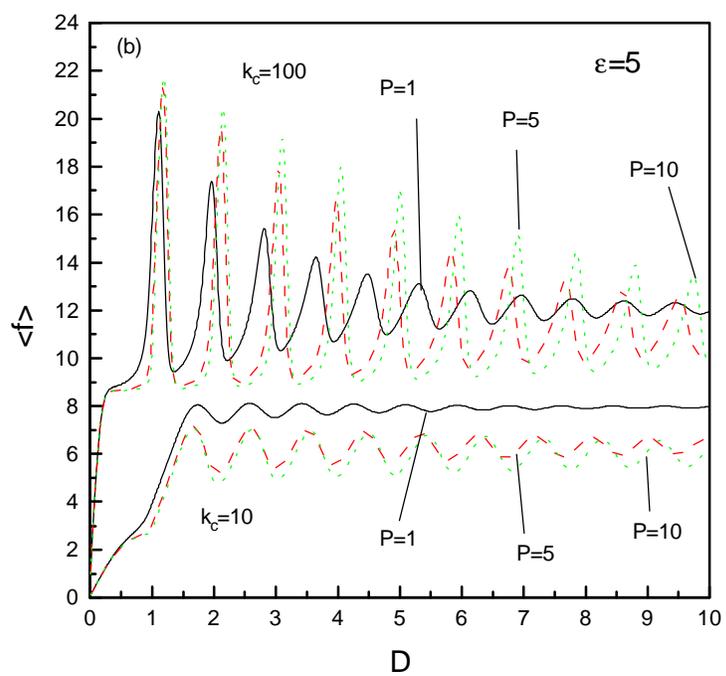

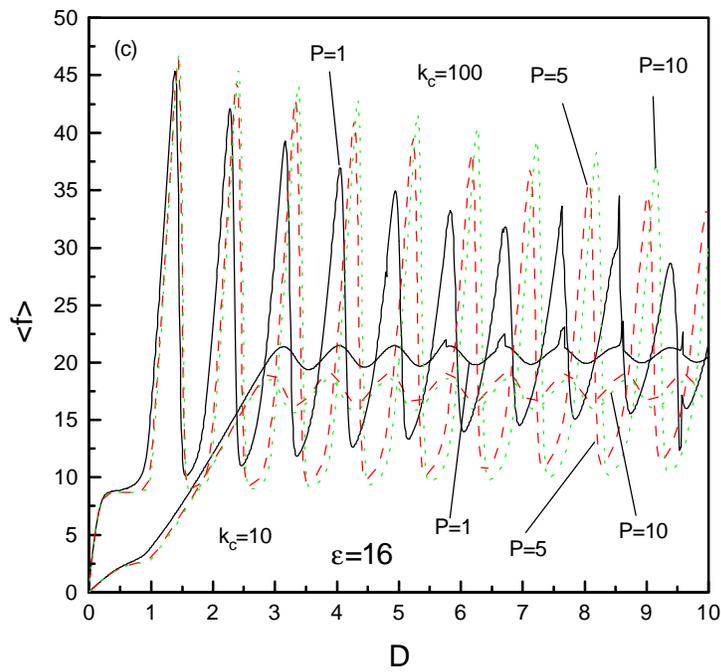

Figure 2: Average force versus cantilever distance D, for two values of the cantilever spring constant $k_c = 10$, and $k_c = 100$ at persistent lengths P=1, P=5 and P=10. The adsorption energies are at (a) $\tilde{e} = 2$, (b) $\tilde{e} = 5$ and (c) $\tilde{e} = 16$.

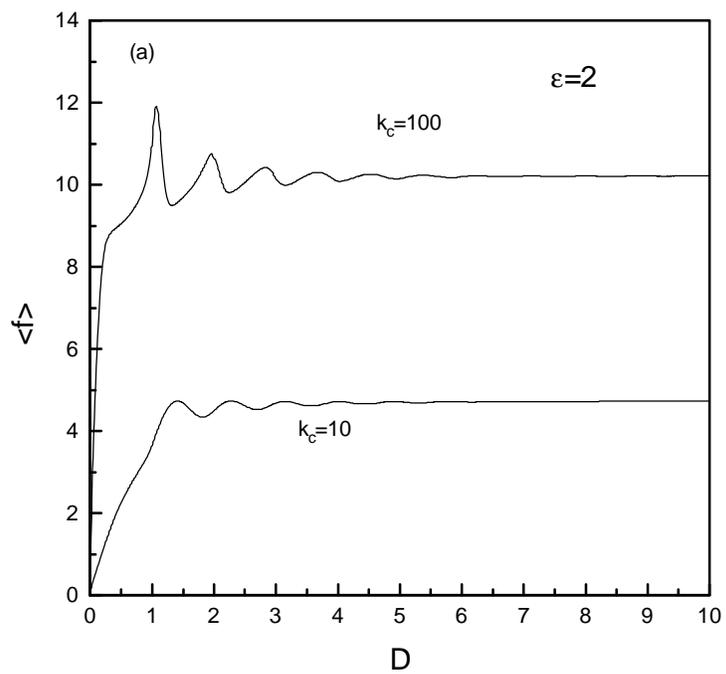

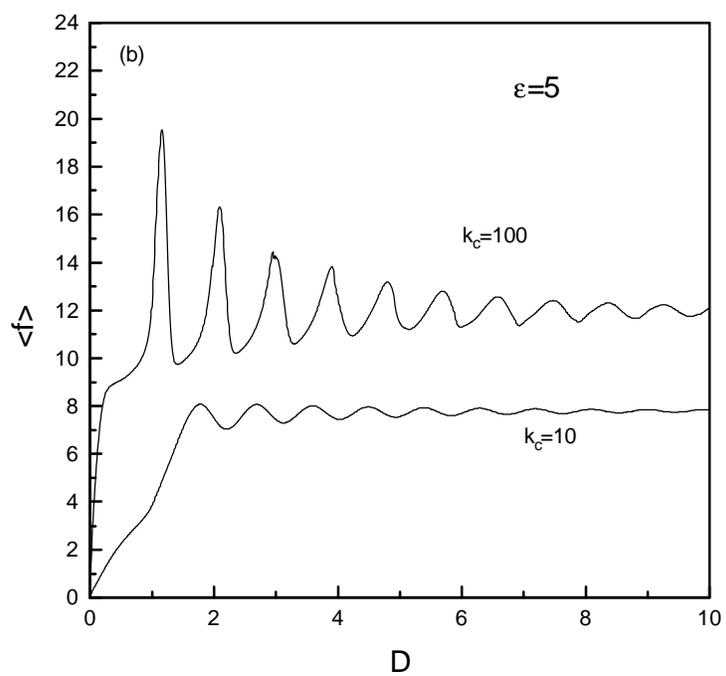

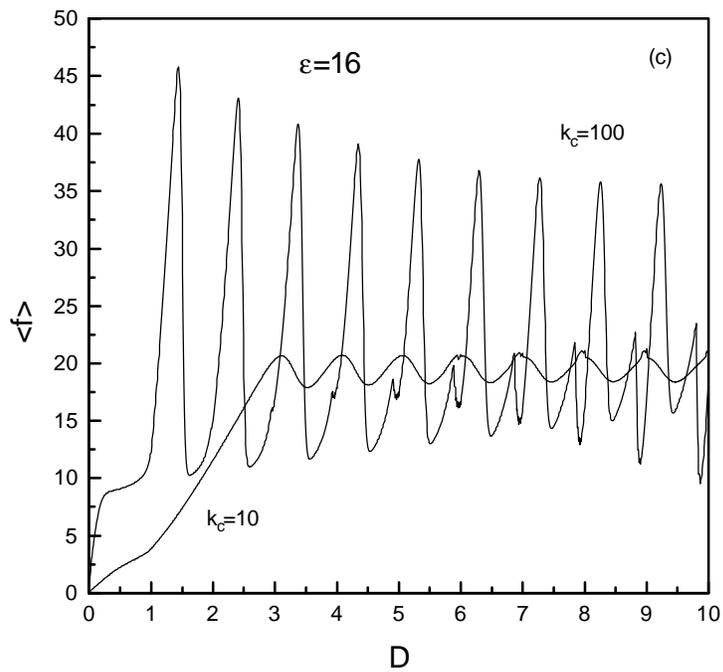

Figure 3: Average force versus cantilever distance D, for two values of the cantilever spring constant $k_c = 10$, and $k_c = 100$, for the FJC model. The adsorption energies are at (a) $\tilde{e} = 2$, (b) $\tilde{e} = 5$ and (c) $\tilde{e} = 16$.